# Ultra-Low-Loss Polaritons in Isotopically Pure Materials: A New Approach

Alexander J. Giles,[1] Siyuan Dai,[2] Igor Vurgaftman,[1] Timothy Hoffman,[3] Song Liu,[3] Lucas Lindsay,[4] Chase T. Ellis,[1] Nathanael Assefa,[5] Ioannis Chatzakis,[6] Thomas L Reinecke,[1] Joseph G. Tischler,[1] Michael M. Fogler,[2] J.H. Edgar,[3] D.N. Basov,[2,7] Joshua D. Caldwell[1]

[1]United States Naval Research Laboratory, Washington, DC
[2]Dept. of Physics, University of California San Diego, San Diego, CA
[3]Dept. of Chemical Engineering, Kansas State University, Manhattan, KS
[4]Materials Science and Technology Division, Oak Ridge National Laboratory, Oak Ridge, TN
[5]NREIP Summer Student residing at NRL, Washington, DC
[6]ASEE Postdoctoral Fellow residing at NRL, Washington, DC
[7]Department of Physics, Columbia University, New York, NY

**Conventional optical components are limited to size-scales much larger than the wavelength of light, as changes to the amplitude, phase and polarization of the electromagnetic fields are accrued gradually along an optical path. However, advances in nanophotonics have produced ultrathin, so-called "flat" optical components that beget abrupt changes in these properties over distances significantly shorter than the free space wavelength.[1-8] While high optical losses still plague many approaches,[9] phonon polariton (PhP) materials have demonstrated long lifetimes for sub-diffractional modes[10-13] in comparison to plasmon-polariton-based nanophotonics. We experimentally observe a three-fold improvement in polariton lifetime through isotopic enrichment of hexagonal boron nitride (hBN). Commensurate increases in the polariton propagation length are demonstrated via direct imaging of polaritonic standing waves by means of infrared nano-optics. Our results provide the foundation for a materials-growth-directed approach towards realizing the loss control necessary for the development of PhP-based nanophotonic devices.**

Due to the long free-space wavelengths in the infrared to terahertz spectral domain, the realization of flat and sub-diffractional-scale optical components promises tremendous advances in imaging, communications, integrated photonics and waveguides. Truly nanoscale photonic elements in the infrared require the use of polaritons – quasiparticles composed of an oscillating charge and a photon – that enable the confinement of light to size-scales well below the diffraction limit.[14,15] In this spectral range, the oscillating charge arises from either free carriers within metals and doped-semiconductors (plasmon polaritons, PPs), or ionic charges on a polar crystal lattice (phonon polaritons, PhPs).[11] Generating these polaritons occurs within materials exhibiting a negative real part of the dielectric function, $Re(\varepsilon) < 0$, which occurs at frequencies below the plasma frequency for PPs or between the transverse (TO) and longitudinal (LO) optic phonons (the Reststrahlen band) for PhPs. As optical phonons exhibit much longer lifetimes than free carriers, the optical losses of PhPs are significantly lower than their PP counterparts. This has been experimentally demonstrated by very large quality factors for sub-diffraction optical antennas.[11-13,16] That said, the fast dispersion of the dielectric function within the Reststrahlen band also results in slow group velocities of the PhP modes and thus limits the propagation length.[10,17] Using isotopically pure materials, we demonstrate significant enhancement in PhP propagation, enabling improvements in waveguides and hyperlenses among other potential applications.[18-21]

Hexagonal boron nitride (hBN) is a polar dielectric material that is particularly well suited for nanophotonic components.[12,17-19,22-24] As a two-dimensional van der Waals crystal, hBN exhibits an extremely large crystalline anisotropy resulting from strong, in-plane covalent bonding of boron and nitrogen atoms and weak, out-of-plane, interlayer van der Waals bonding. This gives rise to two spectrally distinct bands where PhPs can be supported, designated as the lower (LR, ~760-820 cm$^{-1}$) and upper (UR, ~1365-1610 cm$^{-1}$) Reststrahlen bands.[12,23] Further, the strong anisotropy results in a large birefringence, where within the Reststrahlen bands, the dielectric permittivities along orthogonal crystal axes are not only different, but opposite in sign. Such materials are referred to as hyperbolic.[25] In the case of hBN, the in-plane permittivity is isotropic and is defined with a single value, $\varepsilon_t$, which is negative (positive) in the UR (LR), while the out-of-plane component $\varepsilon_z$ is positive (negative). Previous results have shown that the long optical phonon lifetimes enabled quality factors in (naturally abundant) hBN nanostructures that are well in excess of the highest reported values in PP-based systems.[12,24]

The most common point defect and a dominant optic phonon scattering mechanism in hBN is the natural isotope variation of boron: ~80% $^{11}$B and ~20% $^{10}$B ($^{14}$N is 99.6% abundant). Thus a change of only one atomic mass unit represents a ~10% change in the boron mass. Therefore, one could anticipate that large increases in optic phonon lifetime could be realized through isotopic enrichment. Through the use of isotopically enriched boron powder as one of the precursors, hBN crystals with isotope purities as high as $^{11}$B=99.2% and $^{10}$B=98.7% were grown from molten metal solutions (see methods).[26,27] The absolute ratio of the two boron isotopes was quantified using secondary ion mass spectrometry (SIMS) for several samples of different isotopic enrichments, including all samples discussed here. Through comparison of the SIMS-determined ratios with in-plane TO phonon energies extracted from Raman spectroscopy, a relationship was established, allowing the isotopic ratio to be approximated via the Raman shift (see Supplementary Information). Further, an estimate of the phonon lifetime was determined from the TO phonon Raman linewidth. In all, 16 different hBN crystals were characterized with Raman spectroscopy, five of which are shown in **Fig. 1a**. A clear spectral shift from the naturally-occurring hBN TO phonon energy (~1366 cm$^{-1}$) is found with isotopic enrichment, specifically an increase of 27 cm$^{-1}$ to ~1393 cm$^{-1}$ for 98.7% h$^{10}$BN (blue curve, **Fig 1a**) and a decrease of 9 cm$^{-1}$ to 1357 cm$^{-1}$ for 99.2% h$^{11}$BN (red curve). A corresponding spectral shift in the LO phonons and therefore the UR band are also identified, with the band for each material characterized by the broad, highly reflective regions in **Fig. 1b**. In addition, the Raman linewidth narrows significantly with increased isotopic purity, with a minimum linewidth achieved for the hBN sample with the highest enrichments. This minimum linewidth was consistent with a factor of ~3 reduction with respect to the naturally abundant materials (a full table of Raman and SIMS results for all crystals studied are provided in the Supplementary Information). Such a reduction in the Raman linewidth is indicative of a commensurate increase in the optic phonon lifetime.

The overall increase in TO phonon lifetime with isotopic purity can be easily discerned **from Fig. 1c**. Phonon lifetimes are limited by intrinsic three-phonon scattering and by disorder scattering from varying isotopic concentrations. Here we obtained the phonon dispersion relations and the anharmonic interactions between phonons from first principles density functional theory (DFT), with phonon-isotope scattering represented as mass defects. This approach has been shown to give good quantitative results for phonon dispersions (e.g. **Fig. S3** in Supplemental Information) and for thermal transport.[28,29] The results for phonon lifetimes, $\tau_{TO}$ in bulk hBN obtained in this way are shown in **Fig. 1c** (open black symbols) as a function of $^{11}B$ concentration and the experimental lifetimes extracted from the Raman spectra in **Fig. 1a** are provided for comparison (solid red symbols). Good quantitative agreement between theory and experiment is obtained, with the exception of the highest enriched samples ($^{11}B$ concentrations within a few percent of 0% and 100%). For these crystals, the experimentally determined lifetimes appear to saturate near 2 ps, falling short of the ~7.6 ps values predicted for isotopically pure materials. To understand the reason for the suppression of the phonon lifetime, additional SIMS analysis was performed to determine the carbon and oxygen impurity levels in the isotopically enriched samples with respect to the naturally abundant crystals. While quantitative values for the oxygen content could not be ascertained, at least 2-3 orders of magnitude higher carbon concentration was found in the highest enriched samples (see Supplementary Information). Such an increased carbon concentration, especially in the presence of suppressed isotopic scattering, may provide an alternative optic phonon scattering pathway, precluding the ability to observe the predicted 7.6 ps lifetimes. Thus, these results imply that even longer phonon lifetimes, and further reductions in optical losses, should be possible if carbon impurity levels can be suppressed through growth modifications.

Just as the isotopically enriched crystals exhibit shifted phonon frequencies, shifted Reststrahlen bands and increased phonon lifetimes (**Fig. 1**), they possess infrared dielectric functions unique from the naturally abundant material. In an effort to determine these dielectric functions, reflectance spectra were collected from the original hBN crystals in a Bruker Hyperion 2000 FTIR microscope and fit using the commercial WVase software from J.A. Woolam, Inc. The real and imaginary components of the permittivities extracted from these fits are provided in **Fig. 2a** and 2b, respectively, for the UR band of the most highly enriched and naturally abundant hBN samples.

As the optical losses of PhPs are intimately tied to the optic phonon scattering rates, the 3-fold increase in TO phonon lifetime within the enriched samples should result in similar increases in the propagation lengths of the hyperbolic phonon polaritons (HPhPs) they support. Such increases can be clearly identified in the spatial plots of the PhP propagation within ~120 nm thick flakes of the highest isotopically enriched hBN with respect to the similar thickness naturally abundant material (**Fig. 3a-c**). These spatial plots were acquired using s-SNOM techniques described in methods and elsewhere.[4,5,23] HPhP propagation is visualized in these experiments by the interference between the SNOM-tip-launched, hBN-flake-edge-launched and hBN-flake-edge-reflected HPhPs, resulting in the oscillatory patterns seen in the spatial plots. By measuring the periodicity of these oscillations, the HPhP wavelength $\lambda_{HPhP}$, and therefore the compression of the optical fields (in-plane wavevector $q$) can be determined. To accommodate for the spectral shifts in the dielectric functions between each of the hBN materials, the s-SNOM plots presented were collected at frequencies where the magnitude of the real part of the dielectric function was approximately the same $[Re(\varepsilon) \sim -6.25]$. From the qualitative comparisons provided in the corresponding line-scans (**Fig. 3d**) two clear conclusions can be drawn: 1) there is a striking increase in the propagation lengths in the highly enriched samples and 2) additional higher-frequency oscillations are clearly present in the interference patterns observed in the SNOM plots collected from these enriched samples as well.

The in-plane wavevector, which also determines the out-of-plane confinement, is given by $q = 2\pi / \lambda_{HPhP}$. From the fast Fourier transform (FFT) of each linescan, $\lambda_{HPhP}$ can be determined as a function of incident frequency, and thus the magnitudes of $q$ can be extracted (see Supporting Information). The dispersion relationship for the HPhPs can then be determined by plotting $q$ as a function of incident frequency. Comparison of these experimental values with the calculated plots of the HPhP dispersion (**Fig. 4**), make clear that the extracted dielectric functions provide a good estimation of the polaritonic behavior of all three samples, and that these higher-frequency oscillatory components correspond to higher-order HPhPs. In these plots, the false color correspond to the theoretically predicted optical losses, while the symbols designate the experimentally determined values from the s-SNOM measurements. While higher-order modes have been previously reported in three-dimensionally confined hBN nanostructures[12,24] and within naturally abundant flakes of hBN with high resolution s-SNOM scans,[18] in the latter case, those modes were only identified very close to the flake edge decayed within a single oscillatory cycle, therefore any quantification of the damping was extremely difficult. Here, because these higher order modes could be directly discerned in the s-SNOM plots with multiple oscillations, up to three higher-order modes were experimentally identified (triangles in **Fig. 4**), each propagating for multiple oscillations, resulting in associated damping similar to those of the fundamental HPhP modes. The shorter propagation lengths measured for these higher order modes were therefore only the result of the more highly confined, shorter $\lambda_{HPhP}$.

While the s-SNOM plots and linescans provide strong qualitative evidence for increased propagation lengths and reduced HPhP damping, to quantify the improvements a direct comparison must be realized. A relevant figure of merit (FOM) may be defined simply as $Q = Re(q) / Im(q)$, where $Im(q)$ is proportional to the fitted half-width at half-maximum (HWHM) of the spectral linewidths extracted from the FFT of the s-SNOM linescans. Since the enriched samples displayed both tip- and edge-launched modes with center frequencies differing by approximately a factor of 2, we have applied a numerical filter around the frequency of the tip-launched polariton. The tip-launched, rather than edge-launched peak was chosen because it displayed greater consistency from one data set to another and was generally stronger in amplitude. The filtered frequency-space signal was transformed back to real space, corrected for the geometric-decay factor proportional to $x^{-1/2}$, and transformed again into the frequency space. The HWHM was extracted from a Lorentzian fit to the resulting profile, as illustrated for a few cases in the Supplementary Information. To verify the accuracy of the fitted linewidths from the FFT, direct real-space fits of $Im(q)$ to the linescans were determined from s-SNOM plots of naturally abundant hBN flakes and were found to be in general quantitative agreement. A full description of this analysis is provided in the Supplementary Information.

By analyzing the propagating FOM for each of the samples at several incident frequencies, it is clear that there is a substantial improvement in the two isotopically enriched samples (blue squares and red triangles in **Fig. 5a**) over the naturally abundant response (purple circles). The most reliable data points indicate that the improvement ranges from ~50% to more than a factor of 2. While the experimentally observed FOMs are always larger for the enriched samples, the overall magnitude of this increase is less than anticipated based on calculations of the FOM using the extracted dielectric functions. There are many potential contributing factors to this apparent suppression of the FOM and corresponding propagation length, $L_P$ (left axis, **Fig. 5b**). Perhaps the most likely reason stems from the significantly smaller size of the enriched flakes (~10 μm on a side) with respect to the naturally abundant material (>50 μm on a side). This smaller size causes the detected s-SNOM response to consist of two propagating HPhPs, each propagating from opposite edges of the flake that can be observed to meet and thus interfere with one another at the flake center. This causes additional geometrical damping that is not inherent to the material and not present in the larger, natural abundant flakes. Further, this smaller size also made it more difficult to identify flakes of the appropriate thickness free from defects and ridges within a propagation length of the edge, thus resulting in further geometric scattering. Thus, the values of the propagation lengths for the isotopically enriched samples can be considered as a lower limit. Even with these physical limitations, the FOM for the fundamental mode clearly displays the same trend as the Raman linewidths, the fitted damping constants, and the qualitative analysis of the s-SNOM plots.

While the comparison of $L_P$ clearly shows an improvement with isotopic enrichment, we can verify that is the case through a more direct comparison of the measured PhP lifetimes to that of the TO phonons. The PhP lifetime, $\tau_{PhP} = L_P/v_g$, where $v_g$ represents the group velocity and is determined by taking the first derivative of the ω-$q$ dispersion relationships (**Fig. 4**). Since reliable values for $L_P$ were only extracted for the principal mode, we confine the comparison to the first branch of the dispersion relationship. The corresponding $\tau_{PhP}$ plot is given in **Fig 5b** (right axis). An intriguing result is observed in that while $L_P$ is longer for the h$^{10}$BN sample, this is compensated by it's faster $v_g$ (see Supplementary Information) due to the broader Reststrahlen band. This leads to the longer $\tau_{PhP}$ to be found in the h$^{11}$BN material. We found that these $\tau_{PhP}$ values are in good agreement with the Raman determined $\tau_{TO}$ (**Fig. 1c**), suggesting $\tau_{PhP}$ can indeed approach the upper limit imposed by the phonon lifetimes.

Through the isotopic enrichment of hBN, we have experimentally demonstrated three-fold increases in $\tau_{TO}$, and demonstrate that this improvement has resulted in a commensurate increase in both the $\tau_{PhP}$ and $L_P$ over the already low-loss naturally abundant hBN crystals. This was observed despite the presence of elevated levels of other point defects (e.g. carbon impurities). While these enhancements are significant, first-principles calculations predict that over an order of magnitude increase is possible, offering similar improvements in the propagating FOM (dashed green line in **Fig. 5a**). In addition to longer propagation lengths, pronounced higher-order modes were identified and were maintained for many oscillations, which may extend the range of polaritonic applications, for instance in higher transmission, spatial resolution and/or resolving power in hBN-based hyperlensing approaches[18,19] or where fine spatial filtering of broad-band spectral components with high transmission efficiency (see Supplementary Information) is desired. While we have demonstrated the potential of isotopic-enrichment for enhancing phonon lifetimes using hBN, this methodology can be equally applied to a broad range of polar dielectric materials, and is widely applicable over a large spectral range. Further, by coupling this approach with other methodologies that offer longer propagation lengths, such as electromagnetic hybrids,[6,7,10,30] we envision that high-efficiency, multifunctional polaritonic devices can result, offering novel opportunities for the next generation of infrared optical components.

**References**

1   Yu, N. & Capasso, F. Flat optics with designer metasurfaces. *Nature Materials* **13**, 139-150 (2014).

2   Kildishev, A. V., Boltasseva, A. & Shalaev, V. M. Planar Photonics with Metasurfaces. *Science (Wash.)* **339**, 1232009 (2013).
3   Li, P. *et al.* Reversible optical switching of highly confined phonon polaritons with an ultrathin phase-change material. *Nature Materials* **15**, 870-875 (2016).
4   Chen, J. *et al.* Optical nano-imaging of gate-tunable graphene plasmons. *Nature* **487**, 77-81 (2012).
5   Fei, Z. *et al.* Gate-tuning of graphene plasmons revealed by infrared nano-imaging. *Nature* **487**, 82-85 (2012).
6   Dai, S. *et al.* Graphene on hexagonal boron nitride as an tunable hyperbolic metamaterial. *Nature Nanotechnology* **10**, 682-686 (2015).
7   Caldwell, J. D. *et al.* Atomic-scale photonic hybrids for mid-infrared and terahertz nanophotonics. *Nature Nanotechnology* **11**, 9-15, doi:10.1038/nnano.2015.305 (2016).
8   Spann, B. T. *et al.* Photoinduced tunability of the reststrahlen band in 4H-SiC. *Physical Review B* **93**, 085205 (2016).
9   Khurgin, J. B. How to deal with the loss in plasmonics and metamaterials. *Nature Nanotechnology* **10**, 1-5 (2014).
10  Caldwell, J. D. *et al.* Low-Loss, Infrared and Terahertz Nanophotonics with Surface Phonon Polaritons. *Nanophotonics* **4**, 44-68, doi:10.1515/nanoph-2014-0003 (2015).
11  Caldwell, J. D. *et al.* Low-Loss, Extreme Sub-Diffraction Photon Confinement via Silicon Carbide Surface Phonon Polariton Nanopillar Resonators. *Nano Lett.* **13**, 3690-3697 (2013).
12  Caldwell, J. D. *et al.* Sub-diffractional, Volume-confined Polaritons in the Natural Hyperbolic Material Hexagonal Boron Nitride. *Nature Communications* **5**, 5221 (2014).
13  Wang, T., Li, P., Hauer, B., Chigrin, D. N. & Taubner, T. Optical properties of single infrared Resonant Circular Microcavities for Surface Phonon Polaritons. *Nano Lett.* **13**, 5051-5055 (2013).
14  Low, T. *et al.* Polaritons in layered two-dimensional materials. *Nature Materials* **16**, 182-194, doi:10.1038/NMAT4792 (2017).
15  Basov, D. N., Fogler, M. M. & Garcia de Abajo, F. J. Polaritons in van der Waals materials. *Science (Wash.)* **354**, 195-203 (2016).
16  Wang, T. *et al.* Phononic Bowtie Nanoantennas: Controlling Ultra-Narrow-Band Infrared Thermal Radiation at the Nanoscale. *Nano Lett.* **Manuscript in Preparation** (2015).
17  Yoxall, E. *et al.* Direct Observation of Ultraslow Hyperbolic Polariton Propagation with Negative Phase Velocity. *Nature Photonics* **9**, 674-678 (2015).
18  Dai, S. *et al.* Subdiffractional focusing and guiding of polaritonic rays in a natural hyperbolic material. *Nature Communications* **6**, 6963 (2015).
19  Li, P. *et al.* Hyperbolic phonon-polaritons in boron nitride for near-field optical imaging. *Nature Communications* **6**, 7507 (2015).
20  Liu, Z., Lee, H., Xiong, Y., Sun, C. & Zhang, X. Far-field optical hyperlens magnifying sub-diffraction limited objects. *Science* **315**, 1686 (2007).
21  Xiong, Y., Liu, Z. & Zhang, X. A simple design of flat hyperlens for lithography and imaging with half-pitch resolution down to 20 nm. *Applied Physics Letters* **94**, 203108 (2009).
22  Kumar, A., Low, T., Fung, K. H., Avouris, P. & Fang, N. X. Tunable Light-Matter Interaction and the Role of Hyperbolicity in Graphene-hBN system. *Nano Lett.* **15**, 3172-3180 (2015).
23  Dai, S. *et al.* Tunable phonon polaritons in atomically thin van der Waals crystals of boron nitride. *Science (Wash.)* **343**, 1125-1129 (2014).
24  Giles, A. J. *et al.* Imaging of Anomalous Internal Reflections of Hyperbolic Phonon-Polaritons in Hexagonal Boron Nitride. *Nano Lett.* **16**, 3858-3865 (2016).
25  Poddubny, A., Iorsh, I., Belov, P. & Kivshar, Y. Hyperbolic metamaterials. *Nature Photonics* **7**, 948-957 (2013).

26   Kubota, Y., Watanabe, K., Tsuda, O. & Taniguchi, T. Deep Ultraviolet Light-Emitting Hexagonal Boron Nitride Synthesized at Atmospheric Pressure. *Science* **317**, 932-934, doi:10.1126/science.1144216 (2007).

27   Hoffmann, T. B., Zhang, Y., Edgar, J. H. & Gaskill, D. K. Growth of hBN using metallic boron: Isotopically enriched h10BN and h11BN. *MRS Proceedings*, 35-40, doi:10.1557/opl.2014.8 (2014).

28   Lindsay, L., Broido, D. A. & Reinecke, T. L. Ab-initio thermal transport in compound semiconductors. *PhRvB* **87**, 165201 (2013).

29   Lindsay, L., Broido, D. A. & Reinecke, T. L. Phonon-isotope scattering and thermal conductivity in materials with a large isotope effect: A first-principles study. *PhRvB* **88**, 144306 (2013).

30   Caldwell, J. D. & Novoselov, K. S. van der Waals Heterostructures: Mid-Infrared Nanophotonics. *Nature Materials* **14**, 364 (2015).

**Acknowledgements**

A.J.G., C.T.E. acknowledge support from the National Research Council (NRC) and I.C. acknowledges support from the American Society of Engineering (ASEE) NRL Postdoctoral Fellowship Programs. Funding for N.A. was provided through the Naval Research Enterprise Internship Program (NREIP) and is currently an undergraduate student at Rice University in Houston, TX. Funding for J.D.C, I.V., J.G.T. and T.L.R. was provided by the Office of Naval Research and distributed by the Nanoscience Institute at the Naval Research Laboratory. Development of the instrumentation is supported by ARO w911NF-13-1-0210 and AFOSR FA9550-15-0478. D.N.B. is the Moore Investigator in Quantum Materials EPIQS program GBMF4533. The hBN crystal growth at Kansas State University was supported by NSF grant CMMI 1538127.  L. L. acknowledges support from the U. S. Department of Energy, Office of Science, Office of Basic Energy Sciences, Materials Sciences and Engineering Division. Funding for L.L. was provided by UT-Battelle, LLC under Contract No. DE-AC05-00OR22725 with the U.S. Department of Energy. SIMS measurements and analysis was provided by Evans Analytical Group, Inc. as part of a work-for-hire agreement. Authors express their thanks to Dr. Kathy Wahl for use of her Raman microscope.

The United States Government retains and the publisher, by accepting the article for publication, acknowledges that the United States Government retains a non-exclusive, paid-up, irrevocable, world-wide license to publish or reproduce the published form of this manuscript, or allow others to do so, for United States Government purposes. The Department of Energy will provide public access to these results of federally sponsored research in accordance with the DOE Public Access Plan (http://energy.gov/downloads/doe-public-access-plan).

**Author Contributions**

The concept for the experiment was developed by J.D.C., T.L.R. and I.V. All hBN crystals were grown by T.H. and S.L. under the direction of J.E. and provided to J.D.C. through an amazing stroke of good fortune. Exfoliation of hBN flakes was performed by J.D.C. and A.J.G., while AFM characterization was provided by A.J.G. Raman and FTIR analysis was provided by J.D.C., N.A. I.C. C.T.E. and J.G.T. Theoretical calculations of the phonon lifetimes were performed by L.L. and T.L.R., while the code for calculating the dispersion relationship of the HPhPs in hBN was developed by M.F. The FFTs and corresponding lineshape fits were created by I.V. s-SNOM measurements were performed within the lab of D.N.B. by A.J.G. and S.D. All authors discussed results at all stages and participated in the development of the manuscript. J.D.C. and A.J.G. wrote the paper.

**Additional information**

Supplementary information is available in the online version of the paper. Reprints and permissions information is available online at www.nature.com/reprints. Correspondence and requests should be addressed to J.D.C., while requests for hBN materials should be addressed to J.E.

**Competing financial interests**

The authors declare no competing financial interests.

**Methods**

**h-BN Growth**

Hexagonal boron nitride crystals were precipitated from a nickel-chromium flux at atmospheric pressure.

**Crystal growth of hBN with the natural distribution of isotopes**

hBN crystals with the natural distribution of isotopes (20% B-10 and 80% B-11) were produced using a hot-pressed boron nitride ceramic boat, which served as both the container for the metal flux and as the B and N sources. The flux was a mixture of 50 wt% Ni and 50 wt% Cr powders. After loading the crucible, the furnace was evacuated, then filled with $N_2$ and forming gas (5% hydrogen in balance argon) to ~ 850 Torr. The $N_2$ and forming gases continuously flowed through the system during crystal growth with flow rates of 125 sccm and 25 sccm, respectively. The system was heated to 1550 °C for a dwell time of 24 hours. The hBN crystals were formed by cooling at a rate of 1 °C /h to 1500 °C, then quenching the process to room temperature. In our experiments, forming gas was used to minimize oxygen and carbon impurities that are recognized as the main contaminants in hBN crystals.

**$B^{10}$ and $B^{11}$-enriched hBN Crystal Growth Method**

Since hot-pressed boron nitride ceramics are only available with the natural distribution of boron isotopes, to grow isotopically pure hBN crystals, the procedure had to be modified to use elemental boron as a source material. Therefore, the boron nitride boat was replaced with an alumina crucible. High purity $^{10}$B (99.22 at%) or $^{11}$B (99.41 at%) powders were mixed with Ni and Cr powders to give overall concentrations of 4 wt% B, 48 wt% Ni and 48 wt% Cr. In this case, all nitrogen in the hBN originated from the flowing $N_2$ gas. Other than these changes, the procedure was the same as described above.

**Raman Measurements**

Raman measurements were performed using the 532 nm laser line of an argon-ion laser within a Renishaw Raman microscope. The < 10 mW laser line was directed at the sample through a 50x, 0.75 NA objective, with the Raman scattered light being collected back through the same objective. The scattered light was dispersed using a 2400 groove/mm grating onto a silicon CCD.

The spectral position of the Raman lines was calibrated against an internal silicon reference sample. For each different hBN sample studied, 6 measurements were performed with the average spectral positon, linewidth and standard deviations reported (Supplementary Information).

**FTIR Measurements**

Infrared reflectance measurements were performed using a Bruker Hyperion microscope coupled to a Bruker Vertex FTIR spectrometer. The measurements were performed using a 15x Cassegrain objective and were collected with a spectral resolution of 2 cm$^{-1}$. The reflectance spectra reported were all in reference to an aluminum mirror.

**Extracting the Dielectric Function**

The dielectric function was extracted for the three crystals with $^{10}$B 98.8%, $^{11}$B 98.7% and naturally abundant boron isotope concentrations. This was performed using the infrared reflectance spectra collected from the hBN crystals described above. The WVASE software (J.A. Woolam, Inc.) was used to perform the fitting of the reflection spectra and therefore extract an accurate estimation of the infrared dielectric function. A biaxial model was used with two "TO-LO" oscillators chosen, one each for the ordinary and extraordinary components. Starting values were taken from those originally reported in our prior work[12] and from the spectral positions of the TO phonons and corresponding linewidths extracted from the Raman measurements of the same crystals. The best fit parameters are provided in the Supplementary Information.

**Phonon Lifetime Calculations**

Bulk hBN lattice parameters ($a$=2.478Å; $c$=6.425Å) were determined for the *AA'* stacking configuration from electronic structure energy minimization using density functional theory in the local density approximation. The harmonic forces, Born charges and dielectric tensor that govern phonon dispersions were determined from density functional perturbation theory, while the anharmonic forces that govern intrinsic phonon interactions were calculated from perturbations of the electronic structure in large super cells. Transition probabilities for anharmonic three-phonon interactions and phonon-isotope interactions are determined from Fermi's golden rule and summed; inverting this gives the phonon lifetimes.

**s-SNOM Measurements and Analysis**

The s-SNOM measurements were performed by focusing the tunable infrared radiation from a Daylight Solutions MirCAT quantum cascade laser system onto the metallized atomic force microscope (AFM) tip of the NeaSpec s-SNOM instrument. Incident light scatters off of the tip, as well as the edge of the hBN flake, providing the necessary change in momentum to[12,23] launch propagating HPhPs provided the light source is within the Reststrahlen band of the hBN. Due to lack of monochromatic laser sources in the LR, our measurements are limited to the UR. The peak-to-peak distance of the interference fringes observed within the s-SNOM spatial plots is equal to $\lambda_{HPhP}$ and $\lambda_{HPhP}/2$ for the edge- and tip-launched modes, respectively. Plotted in **Fig. 3a-c** are the s-SNOM plots collected at 1510, 1480 and 1480 cm$^{-1}$ for the ~120 nm thick $^{10}$B enriched, naturally abundant, and $^{11}$B enriched flakes, for which $Re(\varepsilon)$ = -6.20, -6.25 and -6.3, respectively. The strong dispersion in this regime requires that we image at (or close to) identical permittivity values, which due to the spectral shift in the optic phonons with changing isotopic mass occur at different frequencies for each of the different samples studied. Furthermore, since the dispersion of the HPhP modes of hBN is strongly thickness dependent, the presented s-SNOM results were all collected from 118-120 nm thick flakes on a 280 nm thick $SiO_2$-on-Si substrate.

**Table of contents figure**

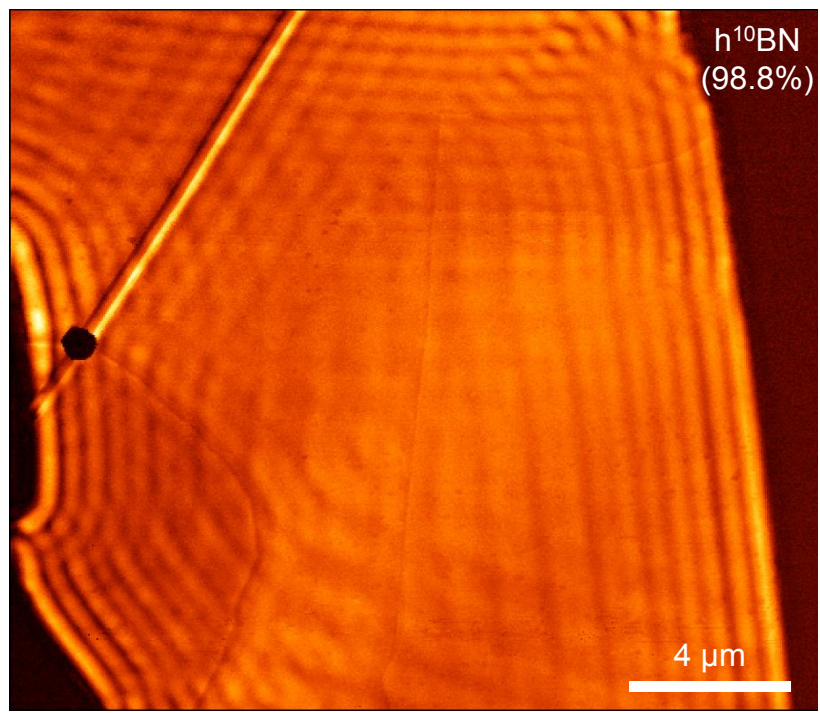

**Figure 1**

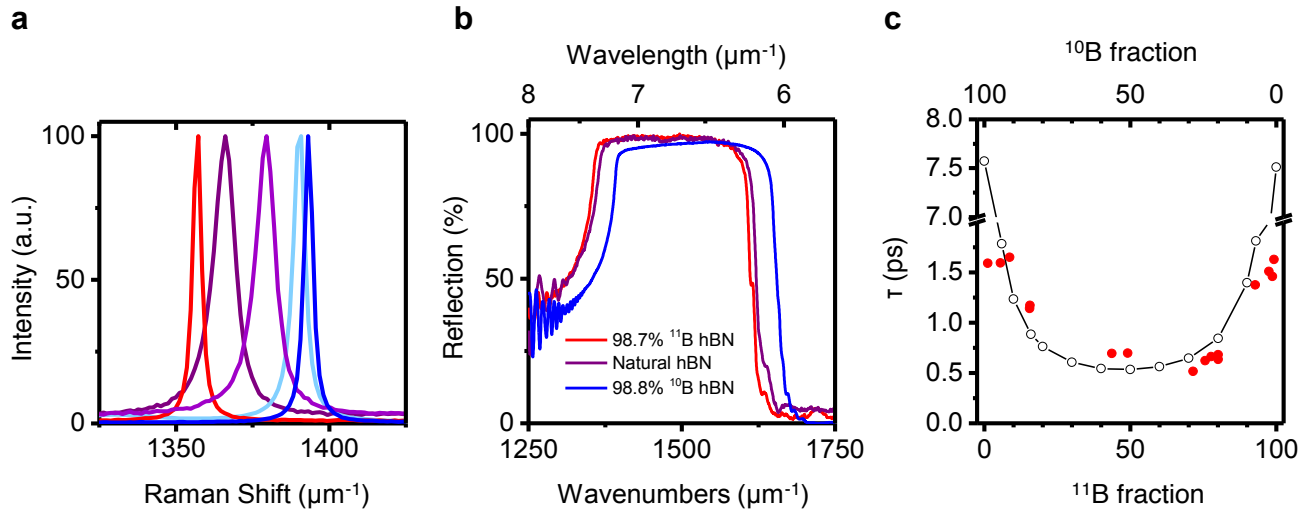

**Fig. 1: Influence of isotopic enrichment on hBN optic phonons. a** 532 nm Raman spectra of hBN crystals with a variety of isotopic enrichments. A clear reduction in linewidth and spectra shift of the TO phonon is observed with the highest enrichments. **b** FTIR reflection spectra of hBN crystals with 98.7% 11B, 98.8% $^{10}$B isotopic purity in comparison to a crystal with the naturally abundant $^{10}$B/$^{11}$B ratio of ~20%/80%. **c** Optic phonon lifetimes for hBN calculated as a function of $^{11}$B enrichment (black open circles; line is a guide to the eye). Lifetimes determined from the Raman measurements in **a** are also plotted (red circles) and show good agreement except at the highest enrichment levels.

**Figure 2**

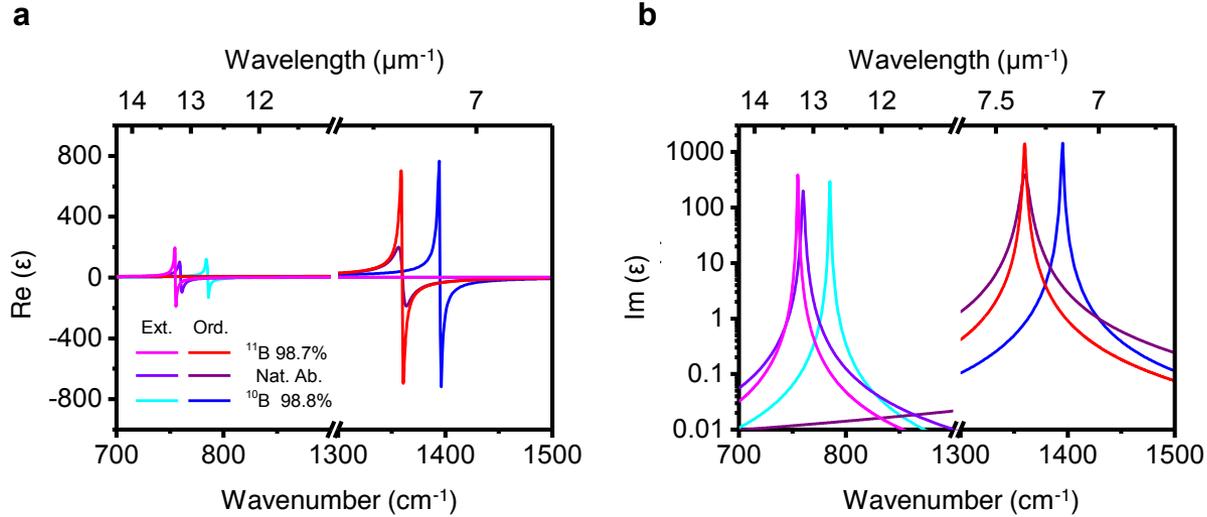

**Fig. 2: Influence of isotopic enrichment on hBN dielectric function. a** Real and **b** imaginary part of the dielectric function for hBN for two highly enriched and for naturally abundant material. The values were extracted from reflectance measurements from hBN crystals that were used to exfoliate the flakes investigated by s-SNOM.

**Figure 3**

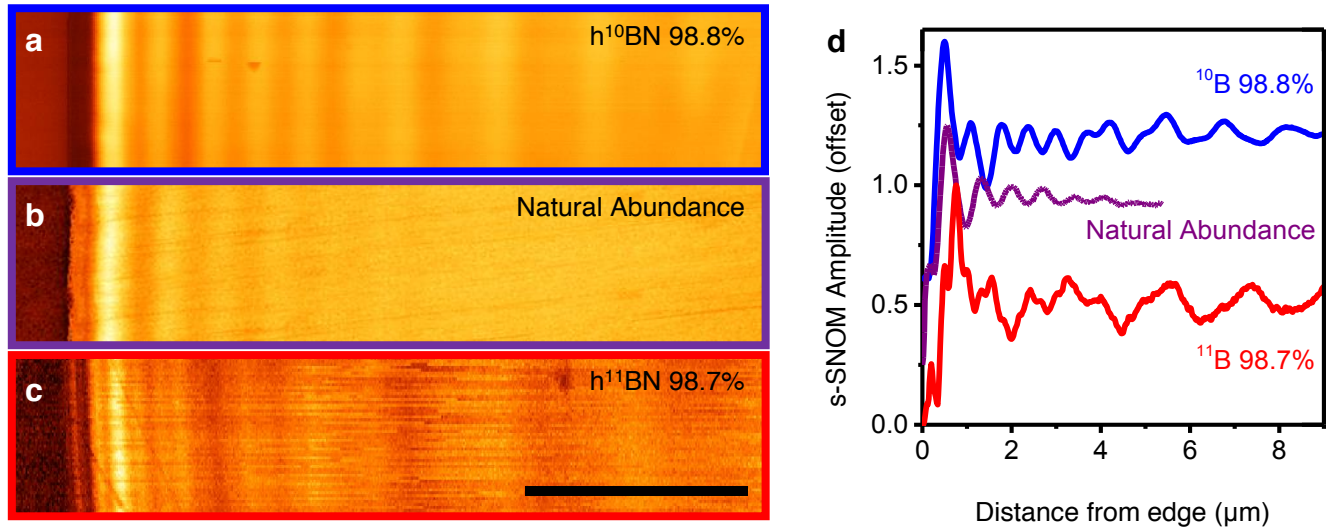

**Fig. 3: Measuring polariton propagation as a function of isotopic enrichment.** Spatial plots of the scattering-type scanning near-field optical microscope (s-SNOM) measurements collected from ~120 nm thick hBN flakes of **a** $^{10}$B 98.8%, **b** naturally abundant and **c** $^{11}$B 98.7%. The scale bar represents 5 μm. The polaritons were stimulated with a 1510, 1480 and 1480 cm$^{-1}$ incident laser source, respectively, corresponding to Re(e)~-6.25 for each. **d** Linescans extracted from **a-c** demonstrate the significantly longer propagation lengths in the enriched in comparison to the naturally abundant flake. The linescans were all normalized and offset so they could be compared on the same scale. Further, the presence of multiple frequency components resulting from higher order modes can be observed in both the raw spatial plots (**a-c**) and the linescans of the enriched samples that are absent from the naturally abundant measurements.

**Figure 4**

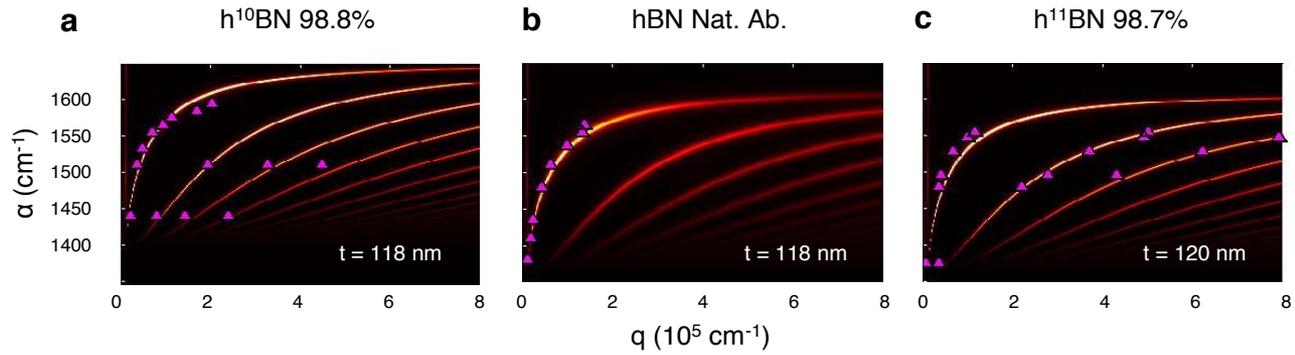

**Fig. 4: Dispersion of hyperbolic phonon polaritons in isotopically enriched hBN.** The in-plane wavevector, $q$, is plotted for the **a** 98.8% $^{10}$B, **b** naturally abundant and **c** 98.7% $^{11}$B hBN flakes in **Fig. 3** as a function of incident frequency. The experimentally extracted fundament and higher order modes (purple triangles) are superimposed upon the calculated dispersion for each using the dielectric functions presented in **Fig. 2**. The loss is qualitatively represented by the breadth and magnitude of the imaginary part of the reflectivity, $r_p$, provided in the false-color plot of the dispersion relation for each crystal type.

**Figure 5**

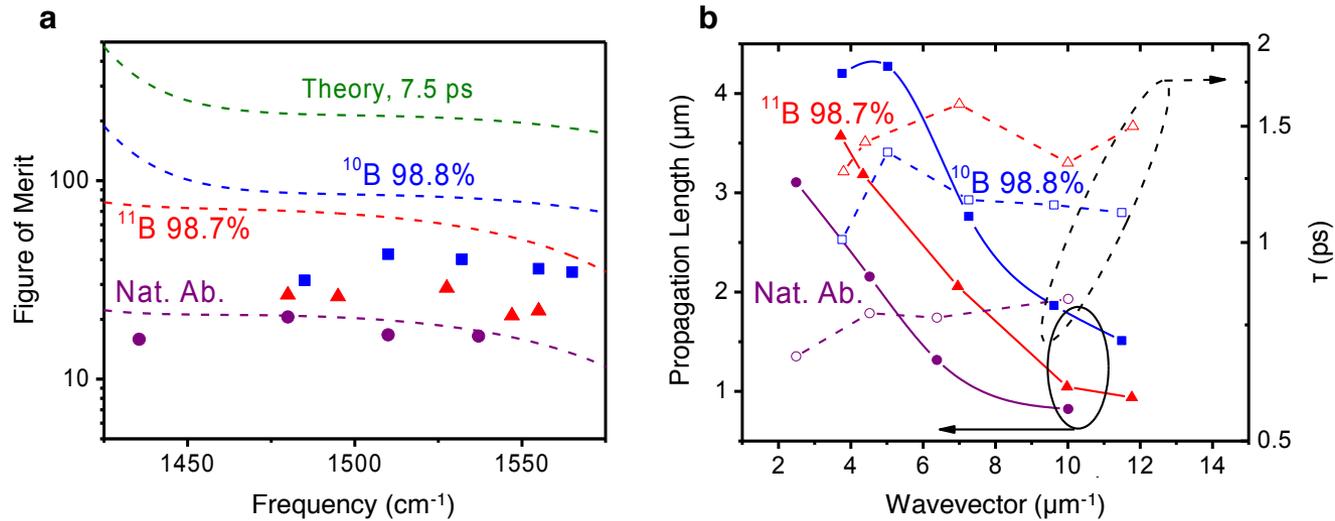

**Fig. 5: Quantifying improvements in loss with isotopic enrichments. a** The figure of merit (FOM) as defined in the text is plotted for each of the hBN crystals presented in **Fig. 3** as a function of incident frequency. The experimental values, extracted from the linewidth and wavevector of each HPhP mode, are plotted for the $^{10}$B 98.8% (blue squares), and $^{11}$B 98.7% (red triangles) enriched hBN flakes in comparison to the naturally abundant sample (purple circles). A clear increase in the FOM is observed for the isotopically enriched samples, however, while the values extracted from the naturally abundant flake agree quite well with the theoretically predicted values calculated from the dielectric function in **Fig. 2** (purple dashed line), the values for the isotopically enriched materials fall short of the corresponding theoretical predictions (red and blue dashed lines). For comparison, the FOM was also calculated for a $^{10}$B with 100% enrichment using the lifetime calculated for such a case in the absence of additional point defects (green line). **b** The increases in FOM also result in commensurate increases in the HPhP propagation lengths (left axis), while extracted HPhP lifetimes for the three samples agree quite well with the phonon lifetimes extracted from the Raman linewidth (**Fig. 1a**).